\documentclass[lettersize,journal]{IEEEtran}

\usepackage{amsmath,amsfonts}
\usepackage{array}
\usepackage[caption=false,font=normalsize,labelfont=sf,textfont=sf]{subfig}
\usepackage{textcomp}
\usepackage{stfloats}
\usepackage{url}
\usepackage{verbatim}
\usepackage{graphicx}
\usepackage{cite}
\usepackage[linesnumbered,ruled,vlined]{algorithm2e}
\usepackage{multirow}
\usepackage{makecell}
\usepackage{tabularx}  % for table
\usepackage{array}  % for table
\usepackage{graphicx}   % for picture
\usepackage{subcaption} % for picture
\usepackage{enumitem}
\usepackage{hyperref}
\usepackage{booktabs}
\usepackage{subfig}
\usepackage{color}
\usepackage{xcolor}
\usepackage{colortbl} % 用于表格颜色
\usepackage{mdframed}
\newcommand{\hash}[1]{\texttt{\small #1}} %hash style
\usepackage{tikz} % ①

\newcommand{\stepnum}[1]{%  % 黑底①
\tikz[baseline=(char.base)]{
\node[shape=circle,fill=black,inner sep=0.1pt] (char)
{\color{white}\textbf{#1}};
}}

\begin{document}

\title{BridgeShield: Risk-Aware Graph Modeling for Cross-Chain Bridge Attack Detection}

\author{Dan~Lin,~\IEEEmembership{Member,~IEEE,}
        Shunfeng~Lu,
        Ziyan~Liu,
        Jiajing~Wu,~\IEEEmembership{Senior Member,~IEEE,}
        Junyuan~Fang,~\IEEEmembership{Member,~IEEE,}
        Jianzhong~Su,
        %Kaixin~Lin,
        Bowen~Song,
        Qing~Xia,
        and~Zibin~Zheng,~\IEEEmembership{Fellow,~IEEE}
    % <-this % stops a space
    \thanks{Received xx March, 2026. The work described in this paper is supported by the National Key Research and Development Program of China (2023YFB2704700), the National Natural Science Foundation of China (62502548, 62332004, 62372485), and is supported by the Ant Group. \textit{ (Corresponding author: Jiajing Wu, Junyuan Fang.)}}
    % <-this % stops a space
    \thanks{Dan Lin, Shunfeng Lu, Ziyan Liu, Jianzhong Su and Zibin Zheng are with the School of Software Engineering, Sun Yat-sen University, Zhuhai 519082, China, also with the GuangDong Engineering Technology Research Center of Blockchain, Zhuhai 519082, China. 
    Jiajing Wu is with Shenzhen Research Institute of Sun Yat-sen University, Shenzhen 518057, China.
    Junyuan Fang is with the Department of Computer Science, City University of Hong Kong, China. 
    %Kaixin Lin is with the School of Computer Science and Engineering Sun Yat-sen University, Guangzhou 510006, China. 
    Qing Xia is with Institute of Software, Chinese Academy of Sciences, Beijing, 100190, China.
    Bowen Song is with Ant Group, Hangzhou 310023, China. }}

% The paper headers
% \markboth{IEEE Transactions on Dependable and Secure Computing,~Vol.~xx, No.~xx, xxx~2026}%
% {Lin  \MakeLowercase{\textit{et al.}}: BridgeShield: Risk-Aware Graph Modeling for Cross-Chain Bridge Attack Detection}

\maketitle

\begin{abstract}
Cross-chain bridges enable asset and state transfers across heterogeneous blockchains, but their complex cross-domain interactions introduce new attack surfaces that are difficult to monitor using traditional single-chain analysis methods. Existing approaches often focus on isolated on-chain behaviors and fail to capture the multi-stage execution semantics of cross-chain transactions. This paper presents BridgeShield, a graph-based framework for detecting cross-chain bridge attacks through risk-aware modeling of cross-chain execution behaviors. BridgeShield reconstructs cross-chain behavior graphs from execution traces and event logs, capturing interactions across the source chain, off-chain relay components, and the destination chain. To highlight attack-relevant structures, the framework employs differential meta-path selection to identify execution patterns that exhibit structural deviations between normal and attack transactions, and hierarchical risk propagation to aggregate distributed risk signals across interaction stages. Experiments on real-world bridge incidents show that BridgeShield achieves an F1-score of 92.6\% in cross-chain attack detection and consistently outperforms existing rule-based and graph-based baselines. In addition, the model remains effective in detecting previously unseen attack incidents, indicating that BridgeShield captures structural risk patterns rather than memorizing historical attack templates.

\begin{IEEEkeywords}
Cross-chain bridges, blockchain security, heterogeneous graph learning, cross-chain attack detection, risk-aware modeling.
\end{IEEEkeywords}

\end{abstract}

\section{Introduction}
\IEEEPARstart{W}{ith} the rapid development of blockchain technology, decentralized finance (DeFi) applications are gradually expanding from a single-chain mode and accelerating towards multi-chain parallel development~\cite{schulte2019towards, belchior2021survey, schar2021decentralized}. Although the multi-chain architecture alleviates the problems of single-chain congestion and high Gas fees, it also brings new challenges: Data and assets on different blockchains cannot flow freely, forming the phenomenon of value silos and hindering the construction of the multi-chain ecosystem. Against this backdrop, cross-chain bridge decentralized applications (DApps) ~\cite{ou2022overview, li2023review} have emerged as solutions to connect these value silos and achieve interoperability between different blockchains.

As cross-chain bridges become a foundational component of decentralized applications, their security has attracted increasing attention from both academia and industry~\cite{zhang2023sok, li2025blockchain}. Due to hidden design vulnerabilities and the substantial value at stake, these bridges have emerged as high-value targets for attackers. According to the Rekt database\footnote{\url{https://rekt.news/zh/leaderboard/}}, three of the five most severe historical blockchain security incidents involved cross-chain bridges, accounting for a cumulative loss exceeding \$1.82 billion. The frequency and severity of such attacks underscore the urgent need to enhance the security of cross-chain bridges. 

Existing studies on blockchain interoperability primarily focus on summarizing known cross-chain bridge vulnerabilities and attacks, while offering potential defense strategies~\cite{haugum2022security, mao2022survey, duan2023attacks, zhao2023comprehensive}. However, most of these efforts lack deployable detection frameworks. Rule-based tools such as XScope~\cite{zhang2022xscope} and XChainWatcher~\cite{augusto2024xchainwatcher} mitigate specific attack patterns to some extent, but their protocol-specific designs severely limit generalizability. More recently, BridgeGuard~\cite{wu2025safeguarding} introduces a graph representation-based detection approach with promising scalability. Nevertheless, it only analyzes transactions confined to a single chain and lacks the capacity to handle off-chain components involved in cross-chain attack execution.
Compared to single-chain DeFi systems, cross-chain bridge applications introduce unique challenges due to their multi-stage execution process spanning multiple blockchains and off-chain coordination components. Two key challenges arise in detecting cross-chain bridge attacks.

\begin{itemize}[leftmargin=*, itemsep=0.6pt]

\item \textbf{C1: Fragmented cross-chain observability.}
A cross-chain behavior spans multiple stages, including a source-chain, off-chain, and destination-chain processing. Execution traces of these stages are distributed across different chains and system layers. Existing detection approaches~\cite{wu2025safeguarding} often analyze transactions within a single chain, making it difficult to reconstruct end-to-end cross-chain interactions. Consequently, attacks that rely on cross-domain coordination may appear benign when observed locally.

\item \textbf{C2: Implicit structural risk patterns.}  
Cross-chain attacks rarely manifest as isolated abnormal transactions. Instead, they often emerge from abnormal combinations of otherwise legitimate operations across multiple entities and execution stages. Such attack behaviors are inherently path-dependent and involve heterogeneous entities including users, bridge contracts, relayers, and token contracts. Conventional detection methods that rely on local features or fixed interaction templates may fail to capture these high-order structural deviations, making it difficult to identify latent risk patterns embedded in cross-chain execution flows.

\end{itemize}

To address these challenges, we propose \textit{BridgeShield}, a graph-based framework for detecting cross-chain bridge attacks through risk-aware modeling of cross-chain execution behaviors. 
First, to mitigate fragmented observability (C1), BridgeShield reconstructs cross-chain execution traces into a unified Cross-chain Behavior Heterogeneous Graph (xBHG), integrating interactions from the source chain, off-chain relay layer, and destination chain into a single representation. 
Second, to capture structural risk patterns (C2), BridgeShield employs differential meta-path selection to identify execution paths that exhibit significant structural deviations between attack and normal transactions. 
Finally, a hierarchical risk propagation mechanism aggregates distributed risk signals across entities, execution patterns, and transactions, enabling robust identification of cross-chain attack behaviors.
The main contributions of this work are summarized as follows:

\begin{itemize}[leftmargin=*, itemsep=0.6pt]

\item \textbf{Problem perspective.}
We study cross-chain bridge attack detection under fragmented observability and implicit structural risks, and formulate it as a unified modeling problem over cross-chain executions spanning the source chain, off-chain relay layer, and destination chain.

\item \textbf{Methodology design.}
We propose \textit{BridgeShield}, a graph-based framework for risk-aware modeling of cross-chain executions. BridgeShield reconstructs cross-chain interaction traces as a heterogeneous behavior graph and introduces differential meta-path selection together with hierarchical risk propagation to capture structural deviations between normal and malicious cross-chain behaviors.

\item \textbf{Empirical evaluation.}
Experiments on our dataset derived from real-world cross-chain bridge incidents demonstrate that BridgeShield effectively detects cross-chain attacks and consistently outperforms existing baseline approaches. The framework achieves an F1-score of 92.6\% and remains capable of identifying previously unseen attack incidents.

\end{itemize}

\section{Related Work}

% 现有研究主要侧重于漏洞分类与防御机制。Augusto 等人、Lee 等人与 Notland 等人分别从多源数据集、架构组件及设计缺陷等维度，系统化地分类了跨链漏洞（如隐私泄露与架构漏洞）并提出预防措施。在检测系统方面，Zhang 等人（Xscope）与 Augusto 等人（XChainWatcher）基于预定义规则与安全属性实现了实时监控，但此类方法通用性较差，难以泛化至新协议。为提升灵活性，Wu 等人提出 BridgeGuard 框架，首次利用图挖掘技术识别新攻击模式。然而，该方法仅侧重于单链交易图分析，忽视了链下攻击场景，难以实现全路径安全监控。
\subsection{Cross-chain Bridge Security.} Existing research focuses on vulnerability taxonomies and defense. Augusto et al.~\cite{augusto2024sok}, Lee et al.~\cite{lee2023sok}, and Notland et al.~\cite{notland2025sok} systematically classified vulnerabilities (e.g., privacy leaks and architectural flaws) based on multi-source datasets and design defects. Related studies on blockchain attack detection, identity alignment, and graph-based analysis~\cite{wu2023defiranger,shi2025invisible,hu2024mfgscope} provide complementary insights, but do not address end-to-end cross-chain attack detection. For cross-chain detection, Xscope~\cite{zhang2022xscope} and XChainWatcher~\cite{augusto2024xchainwatcher} utilized predefined rules and security attributes for real-time monitoring; however, their reliance on specific protocols limits generalization. To enhance flexibility, BridgeGuard~\cite{wu2025safeguarding} introduced graph mining to identify emerging attack patterns, yet its focus on single-chain transactions ignores off-chain scenarios, hindering full-path security monitoring.

% GNN 凭借卓越的拓扑建模能力，已被广泛应用于区块链各领域。在合约漏洞检测上，Zhuang 等人、Liu 等人及 Qin 等人通过构建包含控制流与数据流的复杂合约图（如执行属性图），结合专家知识与时间消息传递网络提升了检测精度。在账户特征分析中，针对分类与钓鱼检测，Liu 等人与 Wang 等人利用特征增强的 GNN 及子图网络实现了高效识别；针对混币溯源难题，Wu 等人（LASC）基于图结构学习实现了对非法资金流的有效关联。尽管图方法成果丰硕，但目前仅有 BridgeGuard 尝试将其应用于跨链桥检测，且其仍存在链下语义缺失的局限。
\subsection{Graph-based Blockchain Analysis.} Graph neural networks (GNNs) are widely used for blockchain security due to their topological modeling capabilities. In vulnerability detection, Zhuang et al. \cite{zhuang2021smart}, Liu et al. \cite{liu2021combining}, and Qin et al. \cite{qin2025enhancing} constructed complex contract graphs (e.g., Execution Property Graphs) combining expert knowledge with temporal networks. For account analysis, Liu et al. \cite{liu2022fa} and Wang et al. \cite{wang2025tsgn} employed feature-enhanced GNNs and subgraph networks for phishing and classification, while LASC \cite{wu2025unveiling} leveraged graph structure learning for mixer deanonymization. Despite these advances, only BridgeGuard \cite{wu2025safeguarding} applies graph modeling to cross-chain bridges, yet it still lacks off-chain semantic integration.

\begin{figure}[tbp]
  \centering
  \includegraphics[width=0.9\linewidth]{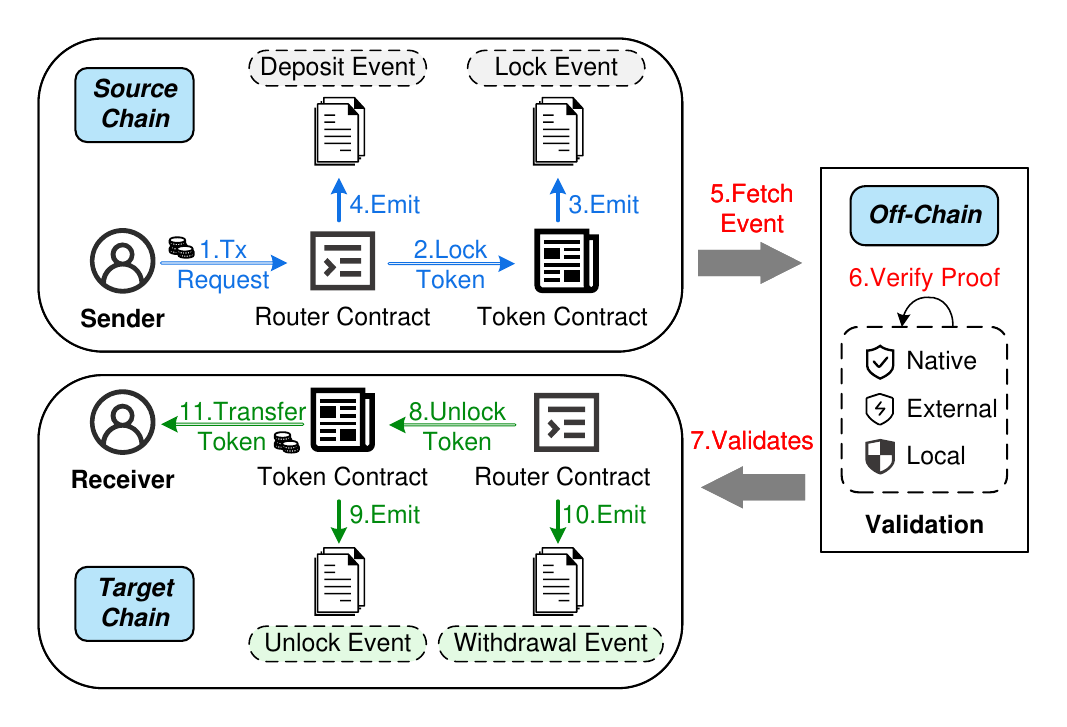}
  \caption{Simplified workflow of a cross-chain bridge.}
  \label{fig01}
\end{figure}

\section{Preliminary}

\subsection{Workflow of Cross-chain Bridges}

% 如图~\ref{fig01} 所示，尽管跨链桥具体实现机制各异，但在本质上均表现为跨越三个阶段的跨域信息同步过程：
% \textbf{源链阶段：} 用户发起请求并触发源链账本的状态变更。根据跨链桥类型的不同，该操作包括：(1) 资产型桥的“锁定/销毁”；(2) 流动性池型桥的“流动性注资”；或 (3) 合成资产型桥的“抵押确认”。
% \item \textbf{链下阶段：} 中继器或预言机捕获源链事件，验证其真实性，并将验证后的意图传播至目标链。
% \item \textbf{目标链阶段：} 目标链合约验证链下证明并执行相应的状态更新,包括“铸造”等价资产、“从本地流动性池兑换”或“发行”合成代币,并将结果分配给用户。
Despite varied implementations, cross-chain bridging fundamentally entails a cross-domain state synchronization process across three stages:
\begin{itemize} [leftmargin=*, itemsep=0.6pt]
\item \textbf{Source chain stage:} A user initiates a request triggering a state change on the source ledger. This involves (1) \textit{Locking/Burning} in asset-mapping bridges, (2) \textit{Liquidity provision} in pool-based bridges, or (3) \textit{Collateralization} in synthetic-asset bridges.
\item \textbf{Off-chain stage:} Relayers or oracles capture source events, verify their validity, and propagate the validated intent to the destination chain.
\item \textbf{Destination chain stage:} The target contract validates the off-chain proof and executes corresponding state updates, such as {minting} assets, {swapping} from local pools, or {issuing} synthetic tokens, before allocating them to the user.
\end{itemize}
Fig.~\ref{fig01} illustrates a simplified workflow of the asset-mapping cross-chain bridge.

\subsection{Problem Formulation}

%A cross-chain attack may unfold across multiple execution stages, including the source chain, the off-chain relay layer, and the destination chain. 
To capture multi-stage behaviors of cross-chain attacks, we model each \emph{cross-chain behavior instance} as an observation unit \(x \in \mathcal{X}\), where \(x\) consists of the execution records associated with one end-to-end cross-chain operation, including source-chain transactions, relay-related messages or traces, destination-chain transactions, event logs, and token transfers.

\begin{mdframed}[linewidth=0.7pt, linecolor=black, skipabove=6pt, skipbelow=6pt, backgroundcolor=gray!5]
\textbf{Problem Definition (Cross-chain Attack Detection).}
\textit{Given a dataset of cross-chain behavior instances \(\mathcal{X}=\{x_1,x_2,\dots,x_n\}\), the goal is to learn a mapping function \(f:\mathcal{X}\rightarrow Y\) that classifies each instance \(x_i\) into one of the following behavior categories:}
\begin{itemize}[leftmargin=*, itemsep=0.6pt]
    \item \textbf{Normal behavior (\(\mathcal{N}\)):} Legitimate cross-chain transactions.
    \item \textbf{Source-chain attacks (\(\mathcal{A}_{src}\)):} Malicious actions on the source chain.
    \item \textbf{Off-chain attacks (\(\mathcal{A}_{off}\)):} Manipulation of relayers, oracles, or validator signatures.
    \item \textbf{Destination-chain attacks (\(\mathcal{A}_{dst}\)):} Malicious execution on the destination chain.
\end{itemize}
\end{mdframed}

% 跨链攻击往往跨越多个执行阶段展开，包括源链、链下中继层以及目标链。为刻画这种多阶段行为，本文将每一次\emph{跨链行为实例}建模为一个观测单元 \(x \in \mathcal{X}\)。其中，\(x\) 表示一次完整跨链操作对应的执行记录集合，包含源链交易、链下中继相关消息或执行痕迹、目标链交易、事件日志以及代币转移等信息。
% \textbf{问题定义（Cross-chain Attack Detection）.}
% \textit{给定跨链行为实例数据集 \(\mathcal{X}=\{x_1,x_2,\dots,x_n\}\)，目标是学习一个映射函数 \(f:\mathcal{X}\rightarrow Y\)，将每个实例 \(x_i\) 分类到以下行为类别之一：}
% \begin{itemize}[leftmargin=*, itemsep=0.6pt]
%     \item \textbf{正常行为 (\(\mathcal{N}\))：} 合法的跨链交易行为。
%     \item \textbf{源链攻击 (\(\mathcal{A}_{src}\))：} 发生在源链上的恶意行为。
%     \item \textbf{链下攻击 (\(\mathcal{A}_{off}\))：} 针对中继、预言机或验证签名的操纵行为。
%     \item \textbf{目标链攻击 (\(\mathcal{A}_{dst}\))：} 发生在目标链上的恶意执行行为。
% \end{itemize}

\section{Attack Incidents and Motivation}
\label{sec:Attack}

This section first describes the collection of real-world cross-chain bridge attack incidents, and then analyzes their execution characteristics to motivate the need for unified cross-chain behavior modeling.

\subsection{Incident Sources}

To obtain representative cross-chain attack cases, we first collect publicly reported bridge security incidents from existing studies and security reports, including Xscope and the BridgeGuard dataset~\cite{wu2025safeguarding}. %In total, 49 bridge-related attack incidents occurring between June 2021 and September 2024 are collected. 
Since this work focuses on attacks that manifest within cross-chain execution processes, we further filter these incidents and retain only those whose attack phases can be observed on the source chain, the relayer layer, or the destination chain. Incidents unrelated to cross-chain execution logic (e.g., rug pull attacks) or lacking sufficient transaction trace information are excluded, thus, 15 incidents remain for further analysis. These incidents cover attacks occurring at different stages of cross-chain execution, including source-chain attacks ($\mathcal{A}_{src}$), relayer attacks ($\mathcal{A}_{off}$), and destination-chain attacks ($\mathcal{A}_{dst}$). 

The statistics of these incidents are summarized in TABLE~\ref{tab:incidents_info}.
For each incident, we further trace the corresponding cross-chain transactions using bridge explorers and on-chain event logs. Specifically, cross-chain transaction instances are identified by correlating the source-chain transaction that triggers the bridge interaction with the corresponding destination-chain transaction that finalizes the cross-chain state transition. These instances serve as the basis for constructing attack transaction samples used in the subsequent dataset construction.

\begin{table}[t]
  \centering
  \caption{Summary of collected bridge attack incidents.}
  \label{tab:incidents_info}
  \footnotesize % 1. 稍微缩小字号，比 scalebox 更标准
  \setlength{\tabcolsep}{3pt} % 2. 极致压缩列间距（默认通常是 6pt）
  \renewcommand{\arraystretch}{1.2}
  \begin{tabularx}{\columnwidth}{lccccX} % 3. 最后一列使用 X 自动换行
    \toprule
    \textbf{Incident} & \textbf{Attack Date} & \textbf{Loss (\$)} & \textbf{\#Tx} & \textbf{ Attack} & \textbf{Bridge} \\
     & & & \textbf{pair} & \textbf{Type} & \textbf{Type} \\ % 4. 表头手动分行
    \midrule
    THORChain \#2   & 2021/07/16  & 7.6M          & 41 & $\mathcal{A}_{src}$   & Liquidity \\
    Qubit           & 2022/01/01   & 80M           & 20 & $\mathcal{A}_{src}$   & Lock-Mint \\
    Meter.io        & 2022/02/06   & 4.3M          & 1  & $\mathcal{A}_{src}$   & Lock-Mint \\
    QAN Platform    & 2022/10/11 & 2M            & 27 & $\mathcal{A}_{src}$   & Lock-Mint \\
    Anyswap \#1     & 2021/07/10  & 7.87M         & 3  & $\mathcal{A}_{off}$ & Liquidity \\
    Levyathan       & 2021/07/30  & 1.5M          & 2  & $\mathcal{A}_{off}$ & Lock-Mint \\
    Ronin \#1       & 2022/03/29  & 610M          & 2  & $\mathcal{A}_{off}$ & Validator \\
    Nomad           & 2022/08/01   & 154M          & 21 & $\mathcal{A}_{off}$ & Lock-Mint \\
    Binance Bridge  & 2022/10/08  & 566M          & 1  & $\mathcal{A}_{off}$ & Lock-Mint \\
    Poly Network \#2 & 2023/07/01   & 10M           & 1  & $\mathcal{A}_{off}$ & Lock-Mint \\
    Ronin \#2       & 2024/08/06   & 12M           & 1  & $\mathcal{A}_{off}$ & Validator \\
    %FEGtoken & 2024/12/29 & 0.9M  & \red{XXX} &$\mathcal{A}_{off}$  & \red{XXX } \\
    ChainSwap       & 2021/07/11  & 4M            & 34 & $\mathcal{A}_{dst}$  & Liquidity \\
    pNetwork        & 2021/09/20  & 1,270M        & 5  & $\mathcal{A}_{dst}$  & Lock-Mint \\
    Ankr            & 2022/12/02  & 5M            & 5  & $\mathcal{A}_{dst}$  & Lock-Mint \\
    Hypr Bridge     & 2023/12/14 & 0.22M         & 1  & $\mathcal{A}_{dst}$  & Lock-Mint \\
    %GemPad   & 2024/12/17 & 1.9M & \red{XXX} &$\mathcal{A}_{dst}$  & \red{XXX } \\
    \bottomrule
  \end{tabularx}
\end{table}

\subsection{Motivation}
% 进一步地，我们对收集到的攻击交易进行细粒度分析，重点考察其在跨链执行过程中的行为特征。
%We conduct a fine-grained analysis of the collected attack transactions.
%, focusing on their behavioral characteristics throughout the cross-chain execution process.
% 部分跨链桥攻击可以在单条链上直接表现出明显的异常行为，例如异常的合约调用序列。然而，我们发现仍有部分的攻击并未在单链上呈现显著异常。
According to the analysis in existing studies~\cite{wu2025safeguarding}, some cross-chain bridge attacks exhibit clear anomalies on a single chain, such as abnormal contract call sequences. However, we observe that a portion of attacks do not present noticeable anomalies when viewed from any individual chain.
% 我们以发生于 2023 年 7 月的 Poly Network #2 事件为例进行分析。在该事件中，攻击者利用链下多重签名验证机制被攻破的问题，通过控制四个多签验证者中的三个，操纵跨链验证流程，从而实现跨链资产的非法提取。值得注意的是，在源链上观测到的存款仅为 1,000 个 LEV 代币，而在目标链上却执行了高达 2,516,430 个 LEV 代币的提现操作。
%We analyze the Poly Network \#2 incident\footnote{\url{https://rekt.news/poly-network-rekt2/}} that occurred in July 2023 as a representative example. The attacker exploited a compromised off-chain multisignature validation mechanism by controlling three out of four multisig validators, thereby manipulating the cross-chain verification process and enabling unauthorized asset withdrawals across chains. Notably, only 1,000 LEV tokens were deposited on the source chain, while a withdrawal of up to 2,516,430 LEV tokens was executed on the destination chain.
We use the \href{https://medium.com/qanplatform/qanx-bridge-disclosure-analysis-continuously-updated-849f7a91d05e}{QAN Platform attack} as a representative example. In this incident, the attacker first deployed a malicious bridge contract on the source chain and submitted a deposit request \href{https://bscscan.com/tx/0x700f1d3a0ef14fb3bd56f97a8a8d034c4e4b79506046f1c7313fc0928f1f35b2}{\hash{0x700f1d...35b2}}
claiming to transfer 4,500,000 QANX tokens. However, due to intentionally modified contract logic, only 1 QANX token was actually transferred. The attacker then obtained signing authorization through deceptive means and initiated a cross-chain withdrawal request \href{https://etherscan.io/tx/0x6bb9fbb8886018524648750606ce7cbe0ec071e9f4c16dc7cdc2cc9a319d8feb}{\hash{0x6bb9fb...8feb}} of 4,802,000 QANX tokens from the legitimate bridge contract on Ethereum.

\begin{figure}[t]
    \centering
    \includegraphics[width=0.8\linewidth]{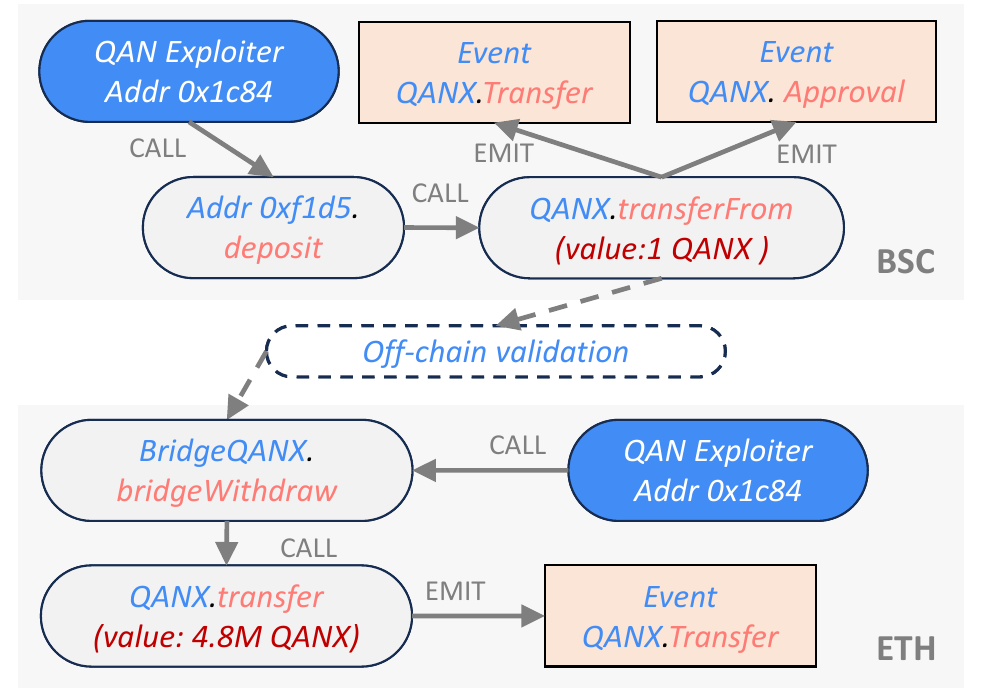}
    \caption{The call graph obtained from the traces of attack transactions of QAN Platform Bridge.}
    \label{fig:qan}
\end{figure}

As shown in Fig.~\ref{fig:qan}, the attack forms a complete cross-chain interaction path spanning the source chain, off-chain components, and the destination chain. A comparison with a normal QAN Platform cross-chain transaction 
shows that the on-chain transactions on both the source and destination chains conform to the expected ABI formats, execution status, and emitted event logs defined by the QAN bridge at that time. Consequently, detection approaches that rely solely on single-chain contract invocation analysis are unable to identify the underlying attack pattern and do not trigger system anomalies or security alerts.

% \footnote{\red{Example transaction hash on BSC chain: }\href{https://bscscan.com/tx/0x59777dc9d6d31910ab5145a0441c6582ce3eb57e38ba28ad64c78fc8571ef9da}{0x5977...f9da}} 
% %0x59777dc9d6d31910ab5145a0441c6582ce3eb57e38ba28ad64c78fc8571ef9da

\begin{mdframed}[linewidth=0.7pt, linecolor=black, skipabove=6pt, skipbelow=6pt, backgroundcolor=gray!5]
\textbf{Finding 1.} 
    Cross-chain attacks often reveal risk only when considering interactions across chains and off-chain coordination. When viewed in isolation on the source or destination chain, their contract call logic closely resembles that of normal transactions.
\end{mdframed}

\section{Detecting Bridge Attacks}
\label{sec:Detection}
This section presents BridgeShield, which models cross-chain behaviors as heterogeneous graphs and detects bridge attacks through risk-aware meta-path modeling and hierarchical risk propagation.

\subsection{BridgeShield Overview}

\begin{figure*}[t]
  \centering
  \includegraphics[width=\linewidth]{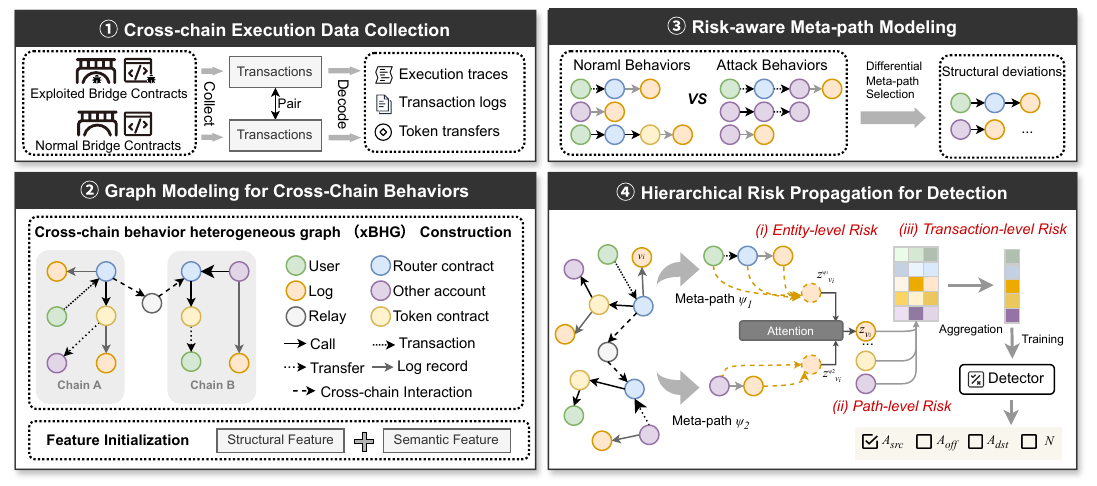}
  \caption{Overview of BridgeShield. Cross-chain execution traces are reconstructed into a heterogeneous behavior graph, where risk-aware meta-path modeling and hierarchical risk propagation are applied to detect cross-chain bridge attacks.}

  \label{fig02}
\end{figure*}

%BridgeShield addresses the challenges of cross-chain attack detection, including fragmented observability across chains and the lack of expressive representations for cross-chain execution semantics. 
As shown in Fig.~\ref{fig02}, the framework consists of four stages.
\stepnum{\small 1} Data collection: We collect cross-chain transactions from exploited and normal bridge contracts, and decodes the execution traces, transaction logs, and token transfers.
\stepnum{\small 2} Graph construction: Based on the decoded data, we construct a cross-chain behavior heterogeneous graph (xBHG) to model entities and their interactions across chains.
\stepnum{\small 3} Meta-path modeling: Deviated interaction patterns are identified structurally between normal and attack behaviors through differential meta-path analysis.
\stepnum{\small 4} Risk propagation: Risk signals are aggregated across entities, paths, and transactions to enable effective detection of cross-chain bridge attacks.

\subsection{Graph Modeling for Cross-Chain Behaviors}

The execution of cross-chain bridges inherently spans the source chain, off-chain relaying components, and the destination chain, with related transactions and logs distributed across different execution domains. 
To capture end-to-end cross-chain behaviors from a global perspective, we model each cross-chain behavior instance as a unified \emph{cross-chain behavior heterogeneous graph}, which organizes execution fragments from two blockchains and their off-chain coordination into a coherent behavioral representation.

\noindent\textbf{Definition (Cross-chain behavior heterogeneous graph). }
\textit{The cross-chain behavior heterogeneous graph (xBHG) is a heterogeneous graph defined as $G=(V,E,\phi,\varphi)$, where $V$ and $E$ denote the sets of entities and interactions involved in cross-chain execution, and $\phi$ and $\varphi$ are the node-type and relation-type mapping functions, respectively. The specific definitions and semantic descriptions of node and edge types are provided in TABLE~\ref{tab:node_types} and~\ref{tab:edge_types}, respectively. }

\begin{figure}[t]
  \centering
  \includegraphics[width=0.85\linewidth]{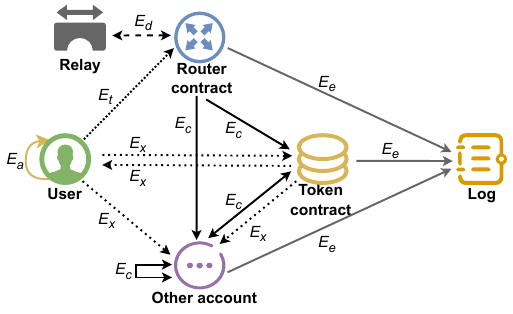}
  \caption{Schema of the cross-chain behavior heterogeneous graph (xBHG).}
  \label{fig03}
\end{figure}

This graph provides a unified abstraction of behavioral information from the source chain, off-chain relaying layer, and destination chain. The graph scheme is shown in Fig.~\ref{fig03}.

\subsubsection{Graph Construction}
\begin{figure}[t]
  \centering
  \includegraphics[width=0.95\linewidth]{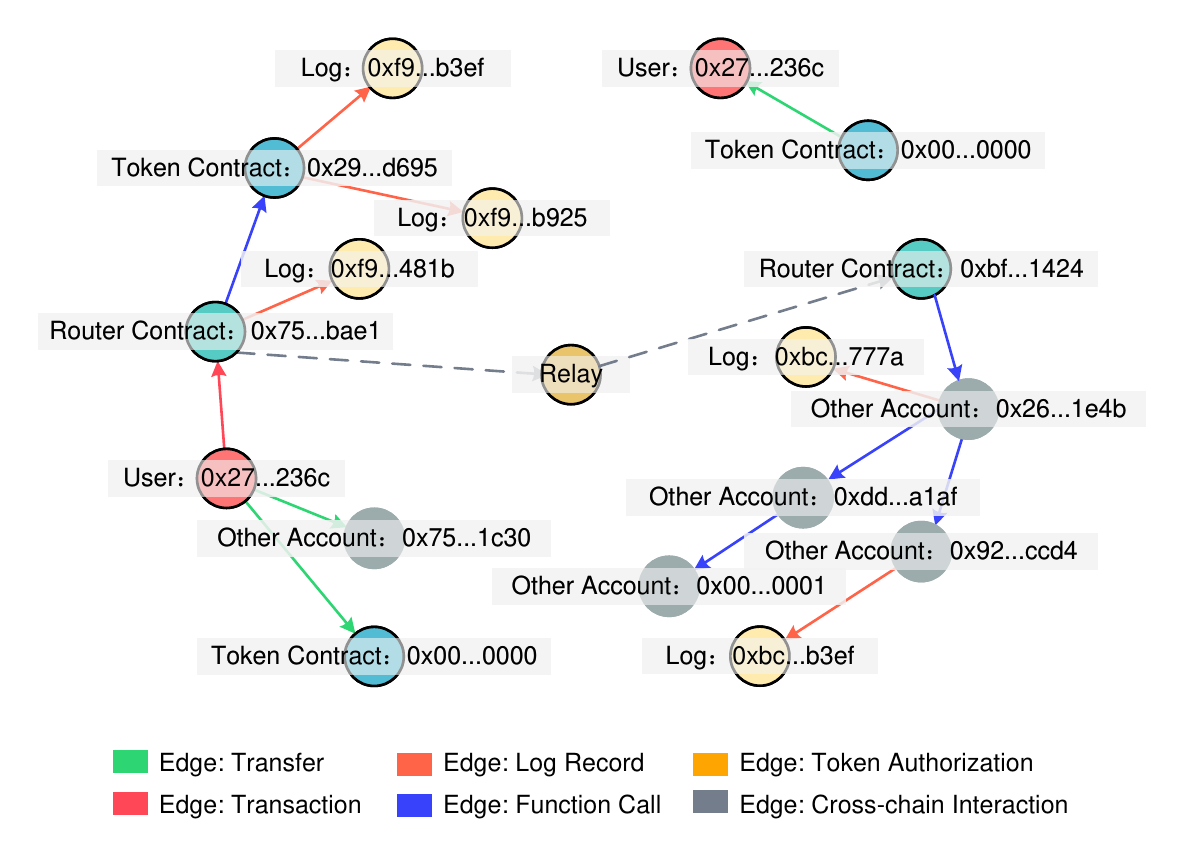}
  \caption{A graph instance constructed from matched transactions between Ethereum and Binance Smart Chain.}

  \label{fig04}
\end{figure}

We extract on-chain behaviors from transaction executions, event logs, and asset transfer records on both the source and destination chains, and introduce off-chain relaying components to model cross-chain message verification and forwarding. Source- and destination-chain transactions are paired using cross-chain linkage information provided by official bridge explorers or indexing services~\cite{lin2025connector}. For example, in a single cross-chain operation, the explorer explicitly associates the source-chain deposit transaction hash with the corresponding destination-chain execution transaction hash, enabling the connection of fragmented on-chain traces into a unified cross-chain execution transaction graph.
An example of the final obtained graph is shown in Fig.~\ref{fig04}, modeling the matched source chain 
transaction \href{https://etherscan.io/tx/0xf9ff957ecca80188d271f10c263078417cace8e184125150de8bedea81a69225}{\hash{0xf9ff95...9225}}
on Ethereum and destination chain transaction \href{https://bscscan.com/tx/0xbc7fafc4fc6335f044c4d74749acf34403201983b476447e9f3329187b64f1cc}{\hash{0xbc7faf...f1cc}} on Binance Smart Chain. The red user node in the source chain initiates a cross-chain transaction request to the cyan router contract node. The router contract then calls the blue token contract to lock assets and generate a log event. Subsequently, it interacts with the destination chain through the purple relay node to realize the transfer of cross-chain assets.

By explicitly modeling the interactions among the source-chain routing contracts, off-chain relayers, and destination-chain routing contracts, the transaction graph enables the characterization of attack patterns that only emerge at the cross-chain composition level.

\begin{table}[t]
\centering
\caption{Entity node type definitions of xBHG.}
\label{tab:node_types}
\renewcommand{\arraystretch}{1.1} % 增加行高，提升换行后的阅读体验
\setlength{\tabcolsep}{6pt}      % 优化列间距

\scalebox{0.9}{
\begin{tabular}{c p{1.8cm} p{6cm}}
\toprule
\textbf{} & \textbf{Node Type} & \textbf{Function} \\
\midrule
$U$ & User Node & Account initiating cross-chain operations \\
$R$ & Router Contract & Bridge contract coordinating cross-chain transactions \\
$T$ & Token Contract & Contract managing token operations \\
$L$ & Log Node & Event log generated during contract execution \\
$D$ & Relay Node & Off-chain relayer forwarding cross-chain messages \\
$O$ & Other Account & Unmarked accounts not in the above categories \\
\bottomrule
\end{tabular}
}
\end{table}

\begin{table}[t]
\caption{Relation edge type definitions of xBHG.}
\label{tab:edge_types}
\renewcommand{\arraystretch}{1.1} % 增加行高，提升换行后的阅读体验
\scalebox{0.9}{
\setlength{\tabcolsep}{6pt}      % 优化列间距
\setlength{\tabcolsep}{6pt}      % 优化列间距
\begin{tabular}{c p{1.8cm} p{6cm}}
\toprule
\textbf{} & \textbf{Edge Type} & \textbf{Function} \\
\midrule
$E_t$ & Transaction & User initiating a cross-chain operation \\
$E_c$ & Function Call & Contract invocation between entities \\
$E_e$ & Log Record & Contract emitting event logs \\
$E_x$ & Transfer & Token transfer between accounts \\
$E_a$ & Authorization & Token approval allowing delegated transfer \\
$E_d$ & Cross chain & Link between on-chain actions and relayer processes \\
\bottomrule
\end{tabular}
}
\end{table}

\subsubsection{Feature Initialization} 
To characterize cross-chain execution roles, we initialize node representations with structural and semantic features and align them into a unified space.

\noindent\textbf{Structural Feature Initialization.}
To capture the structural roles of nodes in the cross-chain behavior graph, we initialize each node with its in-degree and out-degree, which reflect its connectivity patterns along cross-chain execution paths. The in-degree characterizes how frequently a node is referenced by other entities, while the out-degree indicates its level of outward interaction. In cross-chain bridge scenarios, these structural features help distinguish different roles, such as asset initiators, verification components, and receivers, and provide informative structural cues for subsequent behavior modeling~\cite{wahrstatter2023improving}.

\noindent\textbf{Semantic Feature Initialization.}
Some nodes (e.g., called nodes and log nodes) contain function-level textual information, such as function names and parameter lists, which directly encode the business semantics of cross-chain operations. To leverage these semantics, we adopt the pre-trained CodeBERT model~\cite{feng2020codebert}\footnote{\url{https://huggingface.co/microsoft/codebert-base}} to encode function-related texts into 512-dimensional vector representations. Compared with general-purpose language models, CodeBERT is pre-trained on programming language corpora and is thus more suitable for capturing the structured syntax and domain-specific semantics of function-level texts, while maintaining relatively low computational overhead and easy integration with graph learning frameworks.

\noindent\textbf{Feature Alignment.}
As node features vary across different node types in both dimensionality and representation, we apply a unified feature alignment to ensure stable and comparable inputs for subsequent modeling. Specifically, for a node type $t \in T_v$, its features are projected into a unified feature space of dimension $d$ via a linear transformation $h_t = W_t x_t + b_t$,
where $h_t$ represents the projected feature matrix, $x_t$ represents the feature matrix of type $t$.
This alignment module is primarily used for dimensional unification and scale adjustment, while high-order semantics and nonlinear patterns of cross-chain behaviors are modeled by the subsequent heterogeneous graph attention network, thereby preserving expressive power while reducing the risk of overfitting.

% %where $h_t$ represents the projected feature matrix, with a size of $d \times N_t$, that is, the new dimension and the number of elements; $W_t$ is a learnable type conversion matrix, with a size of $d \times d_t$, and is used to map the original features of the elements belonging to type $t$ from $d_t$ dimensions to the new $d$ dimensions; $x_t$ represents the feature matrix of type $t$, with a size of $d_t \times N_t$; $b_t$ is the bias term, with a size of $d \times 1$.
% 接下来的两个表格是使用《不改变意思的前提下，缩短描述的长度》的方法

\subsection{Risk-aware Meta-path Modeling}

After constructing the cross-chain behavior heterogeneous graph, a remaining challenge is how to effectively represent transaction-level subgraph structures so as to capture cross-chain interaction patterns related to security risks. Existing approaches typically rely on graph traversal~\cite{qin2025enhancing} or graph-level representation learning~\cite{wu2025safeguarding}, which are effective for general relational modeling but insufficient to explicitly characterize the multi-stage behavioral semantics arising from cross-chain executions.

As discussed in the Motivation section, attacks on cross-chain bridges often do not originate from a single anomalous node or edge. Instead, they are triggered by abnormal compositions of legitimate operations across different execution domains, such as unexpected correlations among asset locking, off-chain verification, and destination-chain execution. These risk patterns are inherently path-level and compositional, making them difficult to capture using local or homogeneous modeling schemes.

To model such high-level semantics, we adopt meta-paths~\cite{fan2019metapath,fu2020magnn} as semantic abstractions of cross-chain behaviors. In the cross-chain behavior heterogeneous graph, a meta-path consists of an ordered sequence of node and relation types, formally defined as
\[
V_1 \xrightarrow{E_1} V_2 \xrightarrow{E_2} \cdots \xrightarrow{E_n} V_{n+1},
\]
which describes a composite interaction spanning multiple execution domains and functional roles, and thus characterizes key execution structures within transaction-level subgraphs.

However, not all meta-paths are equally informative for attack detection. Here, ``risk-aware'' indicates that meta-paths are selected based on their distributional discrepancies between attack and normal executions, highlighting interaction patterns that exhibit systematic shifts in cross-chain attacks. Accordingly, we perform a differential analysis of meta-path distributions in attack and normal transactions, focusing on those with significant frequency differences.

Specifically, for a given meta-path $\psi_j$, its distributional discrepancy between attack and normal samples is defined as
\begin{equation}
\Delta(\psi_j) = \left| f_{\psi_j}^{\mathcal{A}} - f_{\psi_j}^{\mathcal{N}} \right|,
\label{equ:distributional-meta-path}
\end{equation}
where $f_{\psi_j}^{\mathcal{A}}$ and $f_{\psi_j}^{\mathcal{N}}$ denote the normalized occurrence frequencies of $\psi_j$ in attack and normal samples, respectively. Meta-paths with $\Delta(\psi_j)$ exceeding a threshold $\theta$ are selected as differential meta-paths, as they capture cross-chain interaction patterns indicative of attacks.

Through this risk-aware meta-path modeling, we explicitly extract structurally discriminative semantics from transaction-level subgraphs, which supports generalization to previously unseen attacks. To balance expressiveness and robustness, we restrict the meta-path length to a moderate range (2--4 hops), covering key stages of typical cross-chain executions while avoiding sparsity and reduced interpretability caused by overly long paths.

\subsection{Meta-path Based Risk Propagation on xBHG}

After identifying risk-aware meta-paths, we further characterize how risk propagates along cross-chain executions. Since cross-chain attacks typically arise from interactions among multiple entities and execution stages rather than isolated operations, risk modeling must be conducted hierarchically at the entity, execution-pattern, and transaction levels. To this end, we instantiate risk propagation over meta-paths via attention mechanisms, enabling the model to emphasize security-critical components and execution patterns and to progressively aggregate them into transaction-level risk evidence.

\subsubsection{Entity-level Risk Modeling}
In the cross-chain behavior heterogeneous graph, a meta-path $\psi_k$ characterizes a specific cross-chain execution pattern (e.g., user--routing contract--destination-chain contract). 
However, even within the same execution pattern, different entities contribute unequally to cross-chain risk. Key components, such as routing contracts or relay nodes, directly participate in asset locking, verification, or release, and their behaviors are more likely to reveal potential attack intentions. In contrast, token contracts or log nodes typically execute predefined operations and provide limited signals for risk discrimination. As a result, simply aggregating neighbors under a meta-path using averaging or summation may introduce irrelevant information and weaken the model’s ability to capture critical risk-related signals.

Based on this observation, we introduce an intra-meta-path attention mechanism to adaptively learn the relative importance of entities within the same cross-chain execution pattern. This mechanism assigns different weights to neighboring nodes according to their interaction features, highlighting entities that are more indicative of risk while suppressing noisy or less relevant interactions.

To capture the diversity of cross-chain interactions, we further adopt a multi-head attention structure, enabling the model to attend to different types of entity interactions (e.g., contract invocation versus asset flow) from multiple subspaces. Under a given meta-path, the representation of node $v_i$ is obtained by a weighted aggregation of its neighbors:
\begin{equation}
z^{\psi_k}_{v_i}
= \big\Vert_{m=1}^{M}
\sigma\!\left(
\sum_{v_j \in \mathcal{N}^{\psi_k}_{v_i}}
\alpha^{\psi_k}_{v_i v_j} \cdot h_{v_j}
\right),
\end{equation}
where $\mathcal{N}^{\psi_k}_{v_i}$ denotes the neighbors of $v_i$ under meta-path $\psi_k$, and $\alpha^{\psi_k}_{v_i v_j}$ represents the intra-meta-path attention weight of neighbor $v_j$, reflecting its relative risk contribution under the current execution pattern.

This representation emphasizes entities that are more indicative of risk within the given cross-chain execution pattern while mitigating the influence of irrelevant interactions. The resulting node embeddings $\{z^{\psi_k}_{v_0}, z^{\psi_k}_{v_1}, \ldots, z^{\psi_k}_{v_x}\}$ are then passed to subsequent modules as path-level inputs, enabling fine-grained modeling of entity-level risk contributions within cross-chain executions.

\subsubsection{Execution-pattern Risk Modeling}

Cross-chain attacks rarely rely on a single execution pattern. Instead, they often emerge from abnormal combinations of multiple interaction paths. 
For example, a direct bridge execution
\begin{equation}
\psi_1 = (U \xrightarrow{E_c} R \xrightarrow{E_c} T),
\end{equation}
typically corresponds to standard asset transfer, whereas an indirect forwarding pattern
\begin{equation}
\psi_2 = (U \xrightarrow{E_a} U \xrightarrow{E_c} T),
\end{equation}
may indicate asset relaying through intermediate accounts, which is more suspicious in certain attack scenarios.

As different meta-paths convey distinct security semantics, treating all execution patterns as equally important may obscure those with strong attack implications. Simple averaging or summation implicitly assumes uniform risk contribution across patterns, which is often unrealistic in cross-chain settings.
Motivated by this observation, we introduce inter-meta-path attention to model the relative risk significance of different execution patterns at the transaction level. This mechanism adaptively assigns weights to meta-path representations, highlighting those more indicative of malicious behavior.

Specifically, given entity representations under multiple meta-paths $\{Z_{\psi_1}, Z_{\psi_2}, \ldots, Z_{\psi_P}\}$, inter-meta-path attention computes the corresponding weights $\{\beta_{\psi_1}, \beta_{\psi_2}, \ldots, \beta_{\psi_P}\}$. The final entity representation under transactions is obtained by
% \begin{equation}
$Z_{v_i} = \sum_{k=1}^{P} \beta_{\psi_k} \cdot Z_{\psi_k}.$
% \end{equation}
This design enables the model to focus on execution patterns that carry stronger risk signals, providing a more discriminative foundation for cross-chain attack detection.

\subsubsection{Transaction-level Risk Aggregation}
After modeling risk propagation at the entity and execution-pattern levels, the final step is to aggregate distributed risk signals into transaction-level risk evidence for attack detection. 

To this end, we further aggregate node representations obtained from risk-aware meta-path propagation into a global graph representation. Different aggregation strategies correspond to different assumptions on how risk evidence accumulates. %$\mathcal{Z}$ 
Formally, let $Z_{v_i}$ denote the entity embedding after meta-path-based risk propagation. 
The transaction-level risk representation $\mathcal{Z}$ is defined by aggregation strategies as:
\begin{itemize}[leftmargin=*, itemsep=0.6pt]
\item Mean aggregation ($\mathcal{Z}_{mean} = \frac{1}{N} \sum_{i=1}^{N} Z_{v_i}$) captures global deviations shared across multiple entities and is suitable for modeling widespread anomalous behaviors;

\item  Max aggregation ($\mathcal{Z}_{max} = \max_{i \in [1,N]} Z_{v_i}$) emphasizes salient anomalies from critical nodes, which helps identify attacks dominated by a small number of high-risk components;

\item  Attention-based aggregation ($\mathcal{Z}_{max} = \sum_{i=1}^{N} \alpha_{v_i} Z_{v_i}$) adaptively integrates evidence according to the risk relevance of nodes.
\end{itemize}

Based on the aggregated transaction-level risk evidence, we perform cross-chain attack detection using a standard MLP classifier, which maps $\mathcal{Z}$ to the attack category space and outputs the predicted probabilities of different attack types. This design grounds detection decisions on holistic risk evidence rather than isolated local anomalies, aligning with the fact that cross-chain attacks are typically triggered by interactions among multiple entities and execution stages.

% 因为 Overview 已经有了， summary 有点多余
% \subsection{Summary of BridgeShield}
% BridgeShield detects cross-chain attacks by jointly modeling cross-chain executions and risk propagation. The framework constructs a cross-chain behavior heterogeneous graph (xBHG) to unify execution traces across the source chain, off-chain relay components, and the destination chain, enabling end-to-end representation of cross-chain behaviors. On this basis, risk-aware meta-path modeling identifies execution patterns that exhibit significant distributional differences between normal and attack transactions. Finally, meta-path-based risk propagation hierarchically aggregates risk signals at the entity, execution-pattern, and transaction levels, producing transaction-level risk evidence for attack detection.

\section{Experimental Evaluation}

\subsection{Experimental Setup}

\subsubsection{Dataset}

Based on the incidents described in Section~\ref{sec:Attack}, we further extract transaction-level attack samples from the corresponding execution traces.  
Table~\ref{tab:dataset_statistics} summarizes the statistics of the constructed dataset. 15 incidents cover attacks occurring at different stages of cross-chain execution, including source-chain attacks, relayer-related attacks, and destination-chain attacks. In total, we obtain 165 attack transaction graphs derived from these incidents.
% Based on the selected incidents described in Section~\ref{sec:Attack}, we further extract transaction-level attack samples from the corresponding execution traces. 
% Specifically, the 15 incidents include 4 source-chain attacks, 7 relayer-related attacks, and 4 destination-chain attacks. As summarized in TABLE~\ref{tab:incidents_info}, the 4 source-chain incidents contribute 89 transaction graphs, the 7 relayer-related incidents contribute 45 transaction graphs, and the 4 destination-chain incidents contribute 31 transaction graphs. By aggregating the transactions extracted from these incidents, we construct a dataset containing 165 cross-chain attack transaction graphs.
%
To construct the normal transaction dataset, we collect cross-chain transactions from representative bridge systems that are independent from the attack incidents used in this study. This design avoids potential data leakage between attack and normal samples and prevents the model from learning bridge-specific patterns.
Following the selection strategy adopted in CONNECTOR~\cite{lin2025connector}, we choose bridges that represent the major cross-chain mechanisms~\cite{11424023}, including HTLC-based bridges, notary-based bridges, and relay-based bridges. 
Specifically, we select three widely used bridges: Celer cBridge, Multichain, and Poly Network.
Normal cross-chain transactions from these bridges are collected between September 2020 and April 2023. After data cleaning and normalization, the resulting dataset contains 21,105 normal transaction graphs.

\subsubsection{Implementation details}
The implementation code and dataset are publicly available at \url{https://github.com/Connector-Tool/BridgeShield}.
Experiments are conducted on a server with an Intel Xeon E5-2680 v4 CPU, 30\,GB RAM, and an NVIDIA GTX 1080 Ti GPU. 
The implementation is based on PyTorch 2.0 with CUDA 11.8 under Ubuntu 20.04. 
BridgeShield is trained using the Adam optimizer with a learning rate of 0.001 and cross-entropy loss, where the hidden dimension is set to 128. 
The dataset is evaluated using 5-fold cross-validation to mitigate sampling bias. 
For baseline methods, feature-engineering approaches use the MLPClassifier in Scikit-learn, while graph-based methods adopt node degree as the initial feature representation with a learning rate of 0.01.
% 所有实验在 Intel Xeon E5-2680 v4 CPU、30GB 内存和 NVIDIA GTX 1080 Ti GPU 的服务器上完成。系统环境为 Ubuntu 20.04，模型基于 PyTorch 2.0 与 CUDA 11.8 实现。BridgeShield 使用 Adam 优化器进行训练，学习率为 0.001，损失函数为交叉熵，隐藏维度设为 128。实验采用 5 折交叉验证以降低数据划分带来的偏差。对于基线方法，特征工程方法使用 Scikit-learn 的 MLPClassifier，而图方法以节点度作为初始特征，学习率设为 0.01。

\subsubsection{Evaluation Metrics}
% To comprehensively assess the performance of the proposed model on the cross-chain attack detection task, 
We adopt Precision, Recall, and F1-score as the primary evaluation metrics. Considering the inherent class imbalance in cross-chain attack datasets, we report both class-wise results and overall averaged performance to mitigate bias introduced by any single metric. All experiments are conducted five times, and we report the mean and standard deviation of each metric.

\begin{table}[t]
\centering
\caption{Statistics of the constructed dataset.}
\label{tab:dataset_statistics}
\footnotesize
\renewcommand{\arraystretch}{1.2}
\begin{tabular}{lcc}
\toprule
\textbf{Category} & \textbf{\#Incidents} & \textbf{\#Graphs} \\
\midrule
Source-chain attacks (\(\mathcal{A}_{src}\))      & 4  & 89 \\
Off-chain attacks (\(\mathcal{A}_{off}\))         & 7  & 45 \\
Destination-chain attacks (\(\mathcal{A}_{dst}\))& 4  & 31 \\
Normal behavior instances (\(\mathcal{N}\))            & -- & 21{,}105 \\
% \midrule
% \textbf{Attack total} & \textbf{15} & \textbf{165} \\
\bottomrule
\end{tabular}
\end{table}

\subsection{Overall Detection Performance on Cross-chain Attacks}

\subsubsection{Comparison Methods}
To comprehensively evaluate the effectiveness of BridgeShield in cross-chain attack detection, we categorize the baseline methods based on their security modeling capabilities into the following three groups:

\begin{itemize}[leftmargin=*, itemsep=0.6pt]
    
    \item Domain-specific cross-chain detection methods, which are specifically designed for cross-chain bridge security, representing the most relevant baselines for this task. 
    Xscope~\cite{zhang2022xscope} is a rule-based detection framework that identifies attack behaviors by matching security attributes through key variables and execution patterns.
    BridgeGuard~\cite{wu2025safeguarding} is a graph-based analysis approach that identifies cross-chain attack patterns by mining global and local graph structures.

    \item Structure-aware graph representation learning methods, which focus on learning representations from graph structures or neighborhood relationships but lack explicit modeling of cross-chain execution semantics. 
    GAT \cite{velickovic2017graph} leverages an attention mechanism to assign varying weights to neighboring nodes, facilitating effective representation learning for node classification.
    Graph2vec \cite{narayanan2017graph2vec} is an unsupervised embedding method that encodes the entire graph structure into a vector, capturing global structural features for graph-level tasks.

    \item Relation-aware heterogeneous graph modeling methods, which distinguish between different node and edge types. While they represent state-of-the-art heterogeneous graph learning, their primary objective remains general-purpose representation.
    RGCN \cite{schlichtkrull2018modeling} employs relation-specific convolutional operations to model interactions between diverse node and edge types.
    HGT \cite{yao2020heterogeneous} is a Transformer-based heterogeneous graph model that utilizes self-attention to learn relationships across different entity types.
    HEAT \cite{mo2022multi} is an extension of the HGT architecture that incorporates edge information to further enhance representation capabilities.
\end{itemize}

\begin{table}[t]
\centering
\caption{Experimental results compared with cross-chain attack detection methods.}
\label{tab:security_methods}
\renewcommand{\arraystretch}{1.2}
\begin{tabular}{p{1.2cm} p{0.85cm} c c c}
\toprule
\textbf{Method} & \textbf{Category} & \textbf{Precision (\%)} & \textbf{Recall (\%)} & \textbf{F1-score (\%)} \\
\midrule
\multirow{4}{*}{Xscope}
& $\mathcal{N}$ & 99.63 ± 0.00 & 99.17 ± 0.00 & 99.40 ± 0.00 \\
& $\mathcal{A}_{src}$ & 53.85 ± 0.00 & 100.00 ± 0.00 & 70.00 ± 0.00 \\
& $\mathcal{A}_{off}$ & 100.00 ± 0.00 & 48.70 ± 0.00 & 65.50 ± 0.00 \\
& $\mathcal{A}_{dst}$ & 100.00 ± 0.00 & 52.70 ± 0.00 & 69.06 ± 0.00 \\
\midrule
\multirow{4}{*}{BridgeGuard}
& $\mathcal{N}$ & 99.94 ± 0.07 & 99.98 ± 0.03 & 99.96 ± 0.04 \\
& $\mathcal{A}_{src}$ & 88.14 ± 3.76 & 87.11 ± 3.81 & 87.23 ± 3.59 \\
& $\mathcal{A}_{off}$ & 0.00 ± 0.00 & 0.00 ± 0.00 & 0.00 ± 0.00 \\
& $\mathcal{A}_{dst}$ & 94.56 ± 3.09 & 85.14 ± 4.33 & 90.16 ± 3.44 \\
\midrule
\multirow{4}{*}{BridgeShield}
& $\mathcal{N}$ & 99.98 ± 0.02 & 99.97 ± 0.06 & 99.98 ± 0.03 \\
& $\mathcal{A}_{src}$ & 94.36 ± 7.93 & 93.03 ± 10.35 & 93.53 ± 8.17 \\
& $\mathcal{A}_{off}$ & 89.09 ± 24.39 & 94.17 ± 8.12 & 90.43 ± 17.90 \\
& $\mathcal{A}_{dst}$ & 95.00 ± 11.18 & 93.14 ± 9.60 & 93.78 ± 9.08 \\
\bottomrule
\end{tabular}
\end{table}

\begin{table}[t]
\centering
\caption{Experimental results of structure-aware graph representation learning.}
\label{tab:homo_methods}
\renewcommand{\arraystretch}{1.2}
\begin{tabular}{p{1cm} p{0.85cm} c c c}
\toprule
\textbf{Method} & \textbf{Category} & \textbf{Precision (\%)} & \textbf{Recall (\%)} & \textbf{F1-score (\%)} \\
\midrule

\multirow{4}{*}{GAT}
 & $\mathcal{N}$ & 99.62 $\pm$ 0.40 & 100.00 $\pm$ 0.00 & 99.81 $\pm$ 0.20 \\
 & $\mathcal{A}_{src}$ & 45.72 $\pm$ 42.24 & 48.11 $\pm$ 45.24 & 46.16 $\pm$ 42.27 \\
 & $\mathcal{A}_{off}$ & 20.00 $\pm$ 44.72 & 7.50 $\pm$ 16.77 & 10.91 $\pm$ 24.39 \\
 & $\mathcal{A}_{dst}$ & 31.52 $\pm$ 44.00 & 32.53 $\pm$ 44.65 & 31.67 $\pm$ 43.46 \\

\midrule

\multirow{4}{*}{Graph2vec}
 & $\mathcal{N}$ & 99.45 $\pm$ 0.08 & 99.95 $\pm$ 0.02 & 99.70 $\pm$ 0.04 \\
 & $\mathcal{A}_{src}$ & 40.00 $\pm$ 25.28 & 7.26 $\pm$ 5.03 & 12.09 $\pm$ 8.08 \\
 & $\mathcal{A}_{off}$ & 62.00 $\pm$ 21.68 & 27.83 $\pm$ 16.28 & 35.46 $\pm$ 14.12 \\
 & $\mathcal{A}_{dst}$ & 56.67 $\pm$ 33.50 & 32.92 $\pm$ 13.82 & 40.76 $\pm$ 18.69 \\

\bottomrule
\end{tabular}
\end{table}

\begin{table}[t]
\centering
\caption{Experimental results of relation-aware heterogeneous graph modeling.}
\label{tab:hetero_methods}
\renewcommand{\arraystretch}{1.2}
\begin{tabular}{p{0.8cm} p{0.85cm} c c c}
\toprule
\textbf{Method} & \textbf{Category} & \textbf{Precision (\%)} & \textbf{Recall (\%)} & \textbf{F1-score (\%)} \\
\midrule

\multirow{4}{*}{RGCN}
 & $\mathcal{N}$ & 99.93 $\pm$ 0.06 & 99.26 $\pm$ 1.53 & 99.59 $\pm$ 0.76 \\
 & $\mathcal{A}_{src}$ & 81.96 $\pm$ 40.33 & 98.00 $\pm$ 4.47 & 82.52 $\pm$ 36.21 \\
 & $\mathcal{A}_{off}$ & 45.14 $\pm$ 31.92 & 54.88 $\pm$ 40.64 & 48.42 $\pm$ 34.63 \\
 & $\mathcal{A}_{dst}$ & 91.85 $\pm$ 13.36 & 84.99 $\pm$ 11.45 & 87.98 $\pm$ 10.75 \\

\midrule

\multirow{4}{*}{HGT}
 & $\mathcal{N}$ & 99.98 $\pm$ 0.01 & 99.97 $\pm$ 0.03 & 99.97 $\pm$ 0.02 \\
 & $\mathcal{A}_{src}$ & 91.68 $\pm$ 6.93 & 95.75 $\pm$ 4.29 & 93.56 $\pm$ 4.55 \\
 & $\mathcal{A}_{off}$ & 54.29 $\pm$ 50.91 & 41.79 $\pm$ 40.40 & 46.29 $\pm$ 43.43 \\
 & $\mathcal{A}_{dst}$ & 66.99 $\pm$ 15.66 & 84.94 $\pm$ 9.90 & 74.18 $\pm$ 12.10 \\

\midrule

\multirow{4}{*}{HEAT}
 & $\mathcal{N}$ & 99.90 $\pm$ 0.17 & 99.97 $\pm$ 0.05 & 99.94 $\pm$ 0.08 \\
 & $\mathcal{A}_{src}$ & 60.28 $\pm$ 38.77 & 70.75 $\pm$ 40.04 & 63.69 $\pm$ 37.36 \\
 & $\mathcal{A}_{off}$ & 53.23 $\pm$ 47.60 & 44.05 $\pm$ 33.50 & 46.71 $\pm$ 36.80 \\
 & $\mathcal{A}_{dst}$ & 51.92 $\pm$ 22.54 & 61.71 $\pm$ 38.11 & 50.40 $\pm$ 28.82 \\

\bottomrule
\end{tabular}
\end{table}

\subsubsection{Quantitative Comparison}

We first examine the domain-specific detection methods designed for cross-chain bridges (TABLE \ref{tab:security_methods}). It is observed that Xscope, a rule-driven system, exhibits high recall in specific attack categories, e.g., it achieves 100\% Recall in $\mathcal{A}_{src}$. However, its Precision is limited to 53.85\%, resulting in a moderate F1-score of 70.00\%. This discrepancy suggests that while rule-based heuristics effectively cover known attack patterns, they are susceptible to high false positive rates. 
% Furthermore, their detection stability degrades when attack behaviors involve combinatorial variations across cross-chain execution stages. % 当攻击行为涉及跨链执行阶段的组合变化时，基于规则的检测难以保持稳定性。
Although BridgeGuard achieves improvements through structural modeling—attaining an F1-score of 90.16\% in the $\mathcal{A}_{dst}$ category, it fails to detect off-chain attacks ($\mathcal{A}_{off}$), yielding 0\% for both Precision and Recall. This limitation indicates that relying solely on static execution paths is insufficient to capture the risk signals arising from cross-domain coordination.

Next, we evaluate homogeneous graph representation learning methods that rely solely on structural information (TABLE \ref{tab:homo_methods}). While GAT and Graph2vec achieve F1-scores exceeding 99.7\% in the normal category ($\mathcal{N}$), their performance degrades significantly across attack categories. For instance, GAT yields an F1-score of only 10.91\% on $\mathcal{A}_{off}$, while Graph2vec achieves a mere 7.26\% Recall on $\mathcal{A}_{src}$. These results suggest that cross-chain attacks seldom exhibit prominent structural anomalies from a single-chain perspective and highlight the inherent limitations of homogeneous modeling. By abstracting away the semantics of diverse execution roles, homogeneous models fail to differentiate functional roles within cross-chain processes. Consequently, these models bias toward learning general topological patterns, rendering them insufficient for capturing latent cross-chain attack signals.

Furthermore, relation-aware heterogeneous graph models (TABLE \ref{tab:hetero_methods}) consistently outperform purely structural methods. For instance, HGT achieves an improved F1-score of 93.56\% on $\mathcal{A}_{src}$, suggesting that distinguishing node and relation types helps characterize functional role differences in cross-chain execution. However, such methods remain unstable in the $\mathcal{A}_{off}$ category. For example, HGT yields an F1-score of only 46.29\% with high variance. This indicates that while heterogeneous modeling enhances representation capacity, the lack of explicit risk-path modeling may cause the model to treat off-chain coordination as ordinary interactions, thereby undermining its ability to identify critical risk patterns.

\subsubsection{Analysis of Incorrect Predictions and Security Boundaries}
\label{subsubsec:incorrect_pred}

To better understand the operational scope of BridgeShield, we conduct a case-by-case analysis of the four misclassified attack transactions in the test set and compare them with high-confidence correctly detected samples. The analysis indicates that the misclassifications primarily fall into three categories.
(i) Attack paths structurally similar to legitimate workflows.
Some false negatives exhibit execution topologies that closely resemble normal cross-chain operations, with no substantial deviation in meta-path frequency distributions. In such cases, risk signals are weakly expressed in the structural space, making clear separation difficult. This reflects an inherent limitation of path-differential modeling when dealing with low-frequency, customized attack behaviors.
(ii) Disconnect between structural behavior and asset-level semantics.
BridgeShield models transaction behavior primarily through execution graph structures and does not explicitly capture asset authenticity, external state dependencies, or cross-chain message contents. When the root cause of an attack lies in logical vulnerabilities or off-chain dependencies, the observable structural deviation on-chain may be limited, thereby defining an observability boundary of the approach.
(iii) Semantically ambiguous samples near the decision boundary.
Cross-chain attacks are highly sparse within the overall transaction distribution. For samples close to the decision threshold, malicious behaviors and legitimate operations may partially overlap in the local structural space. As a result, the model may produce conservative predictions while maintaining overall precision.

% \{错误预测分析与安全边界讨论\}
%  为理解 BridgeShield 在实际场景中的适用范围，我们对测试集中 4 笔误判攻击交易进行了逐笔回溯分析，并与高置信度成功样本进行对比。分析结果表明，误判主要集中在以下三类情形。
%  (1) \textbackslash{}textbf\{结构高度贴近合法流程的攻击路径。\}
% 部分漏报样本在执行拓扑上与正常跨链业务高度相似，其元路径频率分布未呈现明显偏移。在此类情况下，风险信号在结构空间中的表达较弱，难以形成显著区分。这反映了基于路径差异建模在面对低频、定制化攻击时的天然局限。
%  (2) \textbackslash{}textbf\{结构行为与资产语义之间的表达断层。\}
% BridgeShield 主要基于执行图结构建模，对资产真实价值、外部状态或跨链消息内容缺乏直接刻画能力。当攻击根源位于逻辑漏洞或外部依赖层面时，其在执行结构上的异常程度可能有限，从而形成可观测性的边界。
%  (3) \textbackslash{}textbf\{决策边界附近的语义模糊样本。\}
% 跨链攻击在整体交易分布中呈现高度稀疏性。对于接近决策阈值的样本，攻击行为与合法业务在局部结构空间中可能存在重叠，从而导致模型在保持整体精度的前提下给出保守预测。

\subsection{Analysis of Risk-aware Modeling Design Choices}

\begin{figure}
\setlength{\abovecaptionskip}{0.2cm}
    \centering
    \includegraphics[width=0.75\linewidth]{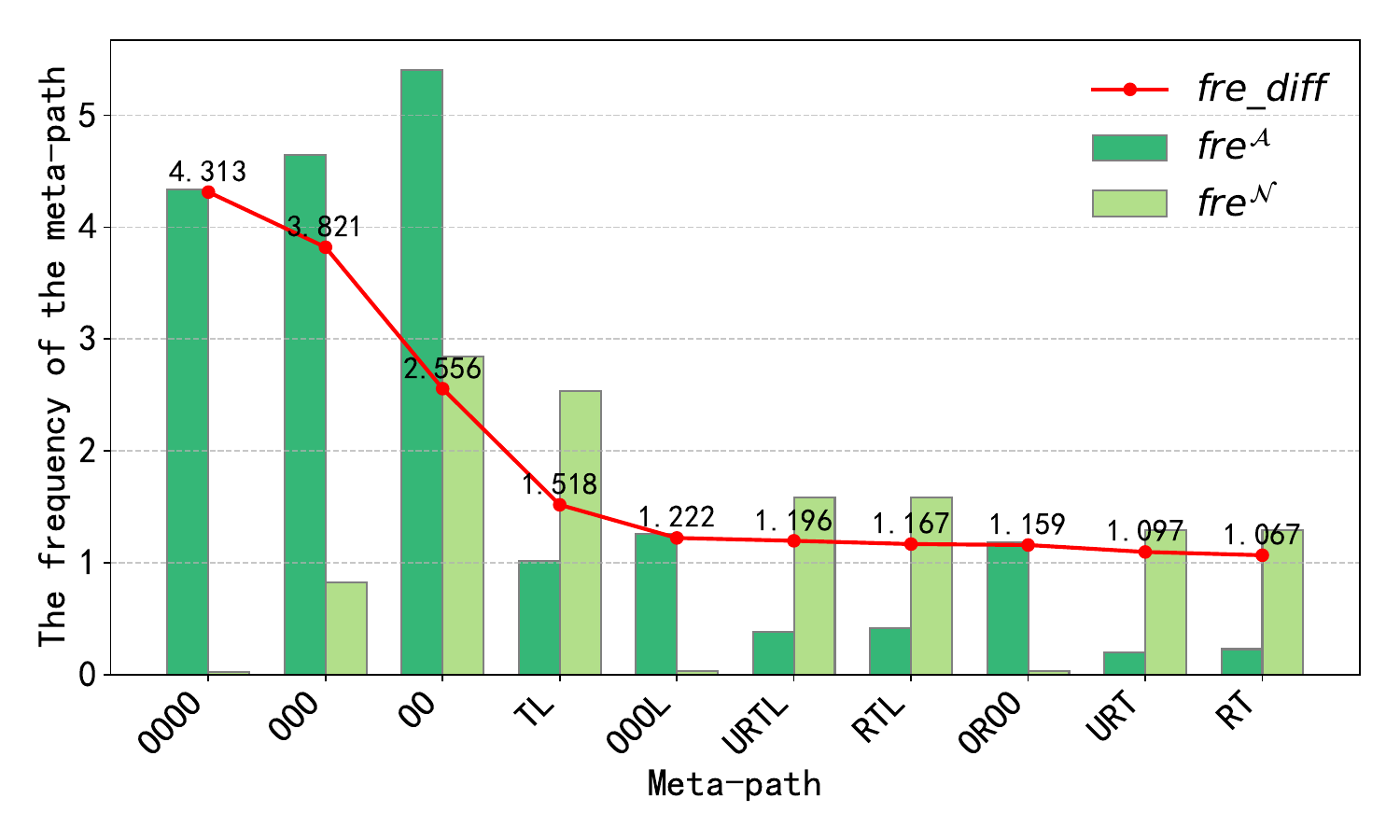}
    \caption{Frequency of meta-paths in normal/attack scenarios.}
    \label{fig06}
\end{figure}

\begin{figure}
\setlength{\abovecaptionskip}{0.2cm}
    \centering
    \includegraphics[width=0.75\linewidth]{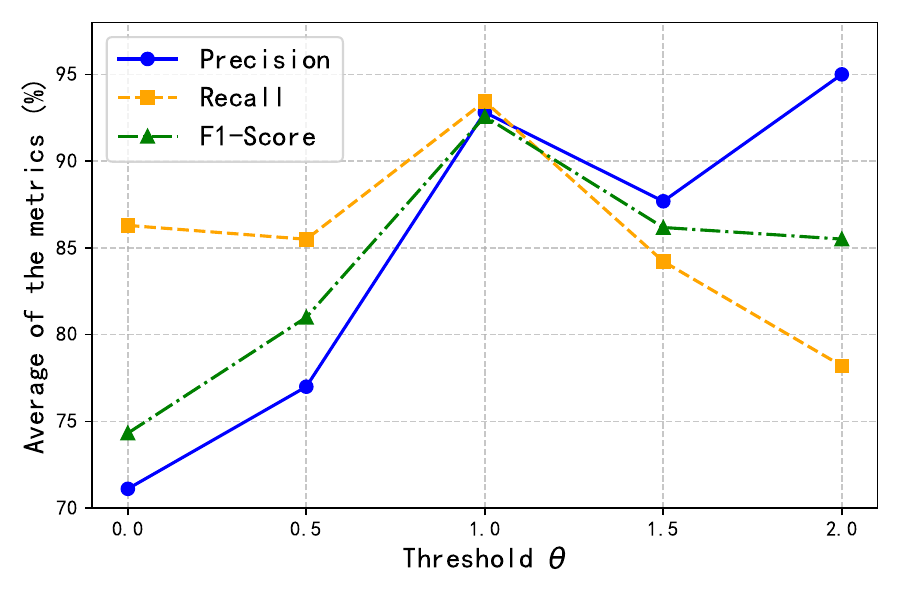}
    \caption{The impact of thresholds on BridgeShield.}
    \label{fig05}
\end{figure}

% \begin{figure}[t]
%     \centering
%     \footnotesize
%     \subfloat[\footnotesize]{ % The frequency of meta-paths under different labels
%         \includegraphics[width=0.45\linewidth]{Figure/06_Frequency_difference_top_ten.pdf}
%         \label{fig05:a}
%     }
%     \hfil
%     \subfloat[\footnotesize]{ % The impact of different thresholds
%         \includegraphics[width=0.45\linewidth]{Figure/05_Different_threshold_results.pdf}
%         \label{fig05:b}
%     }
%     \caption{(a) The impact of thresholds on BridgeShield. (b) The frequency of meta-paths under normal and attack labels when $\theta=1$.}
%     \label{fig05}
% \end{figure}

\begin{table}[t]
\centering
\caption{Effect of risk-aware meta-path selection on transaction-level detection.}
\label{tab:dme_ablation}
\footnotesize
\setlength{\tabcolsep}{4pt}
\renewcommand{\arraystretch}{1.15}

\begin{tabular}{lccc}
\toprule
\textbf{Method} & \textbf{Precision (\%)}  & \textbf{Recall (\%)} & \textbf{F1-score (\%)} \\
\midrule
w/o Meta-paths 
& 71.11 $\pm$ 14.50 
& 86.30 $\pm$ 9.36 
& 74.32 $\pm$ 11.72 \\

BridgeShield (full)                 
% & \textbf{92.82 $\pm$ 0.63} 
% & \textbf{93.45 $\pm$ 0.52} 
% & \textbf{92.58 $\pm$ 0.71}
& 92.82 $\pm$ 0.63 
& 93.45 $\pm$ 0.52 
& 92.58 $\pm$ 0.71
\\
\bottomrule
\end{tabular}
\end{table}

\subsubsection{Effect of Differential Metapath Selection}
% 写作结构
% Effect of Differential Meta-path Selection

% 1. Security semantics of meta-path distributions (Fig a)
% 2. Risk granularity control via θ (Fig b)
% 3. Stability of risk evidence under DME (TABLE X)   ← 新增

This section analyzes the role of the differential metapath selection mechanism in cross-chain attack detection from the perspective of threat modeling. %Unlike conventional parameter sensitivity analysis, the threshold $\theta$ in this study is not employed to tune model complexity. Instead, it serves as a risk semantic constraint designed to identify execution paths that exhibit structural offsets between attack and benign transactions, thereby enhancing the observability of anomalous cross-chain behaviors.

\noindent\textbf{Threat observation.}  %: Role-deficient execution paths
As shown in Fig.~\ref{fig06}, the distribution of metapaths across different labels exhibits significant divergence, reflecting systematic changes in the participant role structure within cross-chain execution processes. In normal business workflows, paths such as ``RTL'' and ``URT'' are more prevalent. These paths typically involve core operational roles, including router and token contracts, which represent the expected cross-chain execution semantics. In contrast, attack transactions more frequently involve paths like ``OOOO'', which primarily consist of unlabeled accounts and lack the participation of core bridge components. This suggests that attackers introduce intermediate accounts or auxiliary contracts to form indirect call chains, bypassing bridge security constraints and reducing anomaly observability. From a security semantics perspective, such ``role-deficient'' execution patterns characterize path obfuscation and trust-boundary evasion, providing the motivation for differential metapath modeling.

\noindent\textbf{Granularity control of risk representation.} 
Fig.~\ref{fig05} shows the performance trends under different thresholds $\theta$. 
When $\theta=0$, all execution paths are retained, introducing numerous low-discriminative patterns that reduce the model’s discriminative ability. 
As $\theta$ increases, the model focuses on risk-relevant execution patterns and reaches optimal performance at $\theta=1$. 
Further increasing $\theta$ leads to a decline in recall, indicating that excessive filtering weakens the coverage of complex attack chains. 
These results suggest that $\theta$ controls the granularity of risk representation by emphasizing structural deviations in cross-chain attacks.
% \noindent\textbf{Granularity control of risk representation.}  % Risk-aware filtering: 
% Fig.~\ref{fig05} illustrates the performance trends of the model under varying thresholds $\theta$. When $\theta=0$, all execution paths are retained. Although this ensures coverage of a broader range of interaction patterns, the inclusion of numerous low-discriminative paths introduces structural noise, which subsequently diminishes the discriminative power of the model. As $\theta$ increases, the model progressively focuses on execution patterns with risk-related semantics and achieves optimal performance at $\theta=1$, indicating that critical risk paths are effectively preserved. When $\theta$ continues to increase, the recall begins to decline, which suggests that excessive filtering weakens the coverage of complex attack chains. This result demonstrates that $\theta$ governs the granularity of risk representation rather than serving as a simple parameter adjustment. Its essential function lies in highlighting the structural offset patterns inherent in cross-chain attacks.

\noindent\textbf{Necessity of Differential Meta-path Selection.}
To evaluate the role of differential meta-path selection in modeling attack structures, we compare the full model with a variant without this module. As shown in Table~\ref{tab:dme_ablation}, removing this module significantly degrades performance, with Precision and F1-score dropping from 92.82\% and 92.58\% to 71.11\% and 74.32\%, respectively, accompanied by noticeably larger variance. Without differential filtering, numerous execution paths shared by normal transactions are incorporated into the modeling process, diluting attack-relevant structural patterns and weakening the model’s ability to distinguish anomalous interactions. In contrast, this module highlights meta-paths that exhibit significant frequency differences between normal and attack behaviors, emphasizing role-absent execution paths and abnormal invocation structures commonly observed in attacks.

\begin{mdframed}[linewidth=0.7pt, linecolor=black, skipabove=6pt, skipbelow=6pt, backgroundcolor=gray!5]
\textbf{Finding 2.} 
    Cross-chain attacks tend to generate execution paths that lack the participation of core roles. Differential metapath selection reinforces structural offset patterns, which stabilizes risk evidence and significantly enhances the observability of anomalous behaviors.
    % 跨链攻击倾向于生成缺乏核心角色参与的执行路径，而差异化元路径筛选通过强化结构性偏移模式，使风险证据更加稳定并显著提升异常行为的可观测性。 
    %Cross-chain attacks tend to form execution paths that lack the participation of core bridge roles and exhibit systematic offsets in metapath structures. Differential metapath filtering effectively captures this risk signal by identifying these structural anomalies.
    % 跨链攻击往往呈现出“角色失衡”的执行模式，其调用路径对桥核心组件的依赖降低，而更多依赖未标记的中间账户，从而在元路径结构上形成可观测的分布偏移。差异化元路径筛选能够有效识别这些风险相关执行模式，而阈值 $\theta$ 实际上控制的是风险表达粒度，通过在噪声抑制与攻击覆盖之间取得平衡，为后续风险传播提供稳定的语义基础。
\end{mdframed}

\subsubsection{Effect of Risk-aware Aggregation Strategies}

\begin{table}[t]
  \centering
  \caption{Performance comparison of different risk-aware aggregation strategies.}
  \label{tab:pooling_methods_results}
  \footnotesize % 缩小字号以适应单栏宽度
  \setlength{\tabcolsep}{1.2pt} % 极致压缩列间距
  \renewcommand{\arraystretch}{1.2}
  
  \begin{tabular}{llccc} % 第1、2列左对齐，后3列等宽居中
    \toprule
    \textbf{Aggregation} & \textbf{Category} & \textbf{Precision (\%)} & \textbf{Recall (\%)} & \textbf{F1-score (\%)} \\
    \midrule
    % 使用嵌套表格模拟垂直居中
    & $\mathcal{N}$       & 99.97 $\pm$ 0.02 & 99.86 $\pm$ 0.12 & 99.91 $\pm$ 0.06 \\
    \textbf{Self-attention} & $\mathcal{A}_{src}$ & 74.02 $\pm$ 18.97 & 93.57 $\pm$ 4.26 & 81.66 $\pm$ 12.50 \\
    \textbf{Pooling}        & $\mathcal{A}_{off}$ & 73.67 $\pm$ 28.29 & 80.50 $\pm$ 19.40 & 71.50 $\pm$ 14.40 \\
    & $\mathcal{A}_{dst}$ & 87.36 $\pm$ 11.80 & 81.90 $\pm$ 26.17 & 80.92 $\pm$ 12.32 \\
    \midrule
    & $\mathcal{N}$       & 99.99 $\pm$ 0.01 & 99.94 $\pm$ 0.10 & 99.96 $\pm$ 0.05 \\
    \textbf{Average}        & $\mathcal{A}_{src}$ & 90.30 $\pm$ 12.42 & 97.03 $\pm$ 4.07 & 93.04 $\pm$ 6.71 \\
    \textbf{Pooling}        & $\mathcal{A}_{off}$ & 81.67 $\pm$ 32.49 & 87.67 $\pm$ 11.64 & 79.59 $\pm$ 23.03 \\ 
    & $\mathcal{A}_{dst}$ & 90.14 $\pm$ 10.53 & 93.14 $\pm$ 9.60 & 91.01 $\pm$ 6.23 \\ 
    \midrule
    & $\mathcal{N}$       & 99.98 $\pm$ 0.02 & 99.97 $\pm$ 0.06 & 99.98 $\pm$ 0.03 \\
    \textbf{Max}            & $\mathcal{A}_{src}$ & 94.36 $\pm$ 7.93 & 93.03 $\pm$ 10.35 & 93.53 $\pm$ 8.17 \\
    \textbf{Pooling}        & $\mathcal{A}_{off}$ & 89.09 $\pm$ 24.39 & 94.17 $\pm$ 8.12 & 90.43 $\pm$ 17.90 \\ 
    & $\mathcal{A}_{dst}$ & 95.00 $\pm$ 11.18 & 93.14 $\pm$ 9.60 & 93.78 $\pm$ 9.08 \\ 
    \bottomrule
  \end{tabular}
\end{table}

\begin{table}[t]
\centering
\caption{Effect of hierarchical risk propagation on transaction-level detection.}
\label{tab:hrp_ablation}
\footnotesize
\setlength{\tabcolsep}{4pt}
\renewcommand{\arraystretch}{1.15}

\begin{tabular}{lccc}
\toprule
\textbf{Method} & \textbf{Precision (\%)} & \textbf{Recall (\%)} & \textbf{F1-score (\%)} \\
\midrule
w/o Risk Propagation 
& 74.57 $\pm$ 1.48 
& 80.23 $\pm$ 1.83
& 75.61 $\pm$ 1.57 \\

BridgeShield (full)                 
% & \textbf{92.82 $\pm$ 0.63} 
% & \textbf{93.45 $\pm$ 0.52} 
% & \textbf{92.58 $\pm$ 0.71}
& 92.82 $\pm$ 0.63
& 93.45 $\pm$ 0.52 
& 92.58 $\pm$ 0.71
\\
\bottomrule
\end{tabular}
\end{table}

This section analyzes the impact of different graph aggregation strategies on cross-chain attack detection from the perspective of risk evidence accumulation. 

Unlike conventional pooling methods in graph learning that prioritize information compression, this study treats aggregation as a process of transforming anomalous signals dispersed across multiple nodes and execution paths into transaction-level risk evidence. Different aggregation mechanisms correspond to distinct risk accumulation hypotheses. Specifically, mean aggregation emphasizes overall behavioral offsets, whereas max aggregation focuses on highlighting intense anomalous signals from a few critical components.

As illustrated in TABLE~\ref{tab:pooling_methods_results}, the average F1-score of self-attention aggregation is only 78.03\%, which suggests that relying solely on learned weights is insufficient for stably capturing critical risk patterns in cross-chain attacks. 
Average aggregation increases the average F1-score to 87.88\%, indicating that shifts in the overall behavioral distribution provide effective risk clues. 
Max aggregation further achieves an average F1-score of 92.58\% and delivers optimal or near-optimal performance across all three attack categories.

These results demonstrate that risk signals in cross-chain attacks are typically characterized by local concentration and uneven intensity. 
Compared with average aggregation which emphasizes global consistency, max aggregation more effectively highlights anomalous patterns triggered by a few high-risk nodes or execution paths. 
This process successfully transforms structural risk clues into more discriminative transaction-level risk evidence. 
The findings also validate the necessity of our risk-aware aggregation design in complex cross-chain scenarios.

% 跨链攻击的风险信号通常呈现为少数关键节点触发的局部异常，而非全局均匀偏移，因此能够保留极端风险贡献的聚合方式更有利于形成稳定的交易级风险证据。
\begin{mdframed}[linewidth=0.7pt, linecolor=black, skipabove=6pt, skipbelow=6pt, backgroundcolor=gray!5]
\textbf{Finding 3.} 
    The risk signals of cross-chain attacks typically manifest as local anomalies triggered by a small number of key nodes rather than uniform global offsets. Consequently, aggregation methods capable of preserving extreme risk contributions are more effective for generating stable transaction-level risk evidence.
\end{mdframed}

\subsubsection{Necessity of Hierarchical Risk Propagation}

Hierarchical risk propagation is designed to characterize the cross-layer accumulation of risk signals within execution paths. %Unlike global aggregation methods that directly process node representations, this mechanism establishes a risk information transmission channel between the path level and the transaction level, thereby preventing local anomalies from being diluted. 
To verify the efficacy of this mechanism, we construct a variant that excludes the hierarchical propagation process. In this variant, global aggregation is performed immediately after obtaining node representations, without any information transmission across path layers. 
The experimental results are illustrated in TABLE~\ref{tab:hrp_ablation}. The findings demonstrate a significant decline in model performance upon the removal of the hierarchical propagation mechanism. This indicates that relying solely on single-layer aggregation is insufficient for identifying structural risks within multi-stage attack chains. Cross-chain attacks often release anomalous signals layer by layer through multi-step calls. In the absence of a hierarchical propagation mechanism, the risk evidence remains fragmented, which weakens the holistic perception of complex attack behaviors.

\begin{mdframed}[linewidth=0.7pt, linecolor=black, skipabove=6pt, skipbelow=6pt, backgroundcolor=gray!5]
\textbf{Finding 4.} 
Risk signals in cross-chain attacks typically accumulate layer by layer along the execution paths rather than existing in isolation.
\end{mdframed}

\begin{table}[b]
\centering
% 设置标题字体稍小，节省垂直空间
%\captionsetup{font=footnotesize, skip=2pt}
\caption{Detailed Transaction Logs of Newly Detected Attacks in the FEGtoken Incident. }
% Hashes are truncated (first 10 + last 6 chars) for layout compatibility.
\label{tab:FEGtoken_attack}

% 使用 tabularx 填满单栏宽度 (\linewidth)
% 列定义：
% l: Case ID (左对齐，固定宽)
% l: Direction (Src/Dst, 左对齐，固定宽)
% X: Chain & Hash (自动填充剩余空间，左对齐)
\renewcommand{\tabularxcolumn}[1]{b{#1}} % 垂直居中对齐
\begin{tabularx}{\linewidth}{@{} l l >{\raggedright\arraybackslash}X @{}}
\toprule
\textbf{Case} & \textbf{Direction} & \textbf{Chain \& Transaction Hash} \\
\midrule

% Case 1
\multirow{2}{*}{\textbf{\#1}} 
& Source & \textbf{BASE}: \href{https://basescan.org/tx/0x5bd9564a732212fa95da2a91ab9a2d9eaef3aedb40b0ad14a4ecee40539d8a94}{\hash{0x5bd956...8a94}} \\
& Destination & \textbf{Ethereum}: \href{https://etherscan.io/tx/0xa0020814a17d45daea9d40654393e7cdebd974ce49a8161958762886a465b89c}{\hash{0xa00208...b89c}}\\
\midrule

% Case 2
\multirow{2}{*}{\textbf{\#2}} 
& Source & \textbf{BASE}: \href{https://basescan.org/tx/0xfc27cf394f01c847fc20f9a162cd07d7e98942260ded3b5557999c8ec1361ea0}{\hash{0xfc27cf...1ea0}}\\
& Destination & \textbf{Ethereum}: \href{https://etherscan.io/tx/0x310ed692dad0dc977ebf4eef90f520eafa262ef67924756b459fb30bfae909c6}{\hash{0x310ed6...09c6}}\\
\midrule

% Case 3
\multirow{2}{*}{\textbf{\#3}} 
& Source & \textbf{BASE}: \href{https://basescan.org/tx/0x355681c1a321ac4ee0537e7fdbf1560c55ecf7eef00659f4cd0addd01d58abb3}{\hash{0x355681...abb3}}\\
& Destination & \textbf{Ethereum}: \href{https://etherscan.io/tx/0x3351a1676afa37fcdd7c61a73f5a93e51b875e1b15174962209b2d73ae271ec0}{\hash{0x3351a1...1ec0}}\\

\bottomrule
\end{tabularx}

\end{table}

\subsection{Generalizability to Unseen Cross-Chain Attacks}

To assess whether BridgeShield generalizes beyond the observed attack distribution, 
we conduct a strict temporal hold-out evaluation using the real-world FEG cross-chain bridge incident. 
The model is trained on data collected prior to time $T_0$, 
while the FEG incident occurred after $T_0$ and was excluded from all training and hyperparameter tuning processes.
The FEG incident was caused by insufficient validation of cross-chain message source addresses, 
which enabled adversaries to trigger unauthorized transfers via forged messages. 
This attack type does not appear in the training set. 
Nevertheless, BridgeShield identifies three cross-chain attack transaction pairs 
within the corresponding time window, as shown in TABLE~\ref{tab:FEGtoken_attack}. 
All detected pairs are assigned risk scores above the decision threshold, 
and no additional transactions are flagged during the same period.

Under default configurations, rule-based and structure-only baselines do not flag these transactions, 
indicating that the detection does not rely on attack-specific signatures or predefined rules. 
Instead, the detected transactions exhibit role-absent execution paths and cross-stage risk accumulation patterns, 
which align with the structural deviations characterized in Section~\ref{sec:Detection}. 
By combining differential meta-path selection with hierarchical risk propagation, 
BridgeShield captures these structural risk constraints, 
suggesting that the model generalizes to unseen cross-chain attack mechanisms 
rather than memorizing historical attack templates.

\section{Conclusion}
This paper presents BridgeShield, a graph-based framework for detecting cross-chain bridge attacks through risk-aware modeling of cross-chain execution semantics. 
By reconstructing cross-chain behavior graphs from execution traces and event logs, BridgeShield captures structural interactions spanning the source chain, off-chain relay components, and the destination chain. 
Differential meta-path selection highlights execution patterns that deviate from normal cross-chain workflows, while hierarchical risk propagation aggregates distributed risk signals across interaction stages. 
Experiments on a dataset of real-world bridge incidents demonstrate that BridgeShield achieves strong detection performance and remains effective under unseen attack scenarios. 
Further analysis shows that the model identifies structural risk patterns rather than relying on memorized attack templates. 
These results suggest that modeling cross-chain interactions as heterogeneous execution graphs provides a practical foundation for generalizable cross-chain bridge security monitoring.

\bibliographystyle{IEEEtran}
\bibliography{BridgeShield_TDSC}

\end{document}